\documentclass[twocolumn]{aastex631}

\shorttitle{Coherent Structures Survey from Early IMAP Observations}
\shortauthors{Gautam et al.}
\graphicspath{{./}{figures/}}
\usepackage{hyperref}

\begin{document}

\title{Scale-Dependent Statistics of Coherent Magnetic Structures in Early IMAP Observations}

\author[0000-0001-7379-4268]{Sujan Prasad Gautam}
\affiliation{Department of Physics and Astronomy, University of Delaware, Newark, DE 19716, USA}

\author[0000-0003-4168-590X]{Francesco Pecora}
\affiliation{Department of Physics and Astronomy, University of Delaware, Newark, DE 19716, USA}
\affiliation{INAF-Istituto di Astrofisica e Planetologia Spaziali, Via del Fosso del Cavaliere 100, Roma, Italy}

\author[0000-0003-2160-7066]{Subash Adhikari}
\affiliation{Department of Physics and Astronomy, University of Delaware, Newark, DE 19716, USA}

\author[0000-0002-
6962-0959]{Riddhi Bandyopadhyay}
\affiliation{Department of Physics and Astronomy, University of Delaware, Newark, DE 19716, USA}

\author[0009-0008-8723-610X]{Jiaming Wang}
\affiliation{Department of Physics and Astronomy, University of Delaware, Newark, DE 19716, USA}

\author[0000-0002-5354-1164]{Joshua Goodwill}
\affiliation{Department of Physics and Astronomy, University of Delaware, Newark, DE 19716, USA}

\author[0009-0004-8086-3013]{William Taranto}
\affiliation{Department of Physics and Astronomy, University of Delaware, Newark, DE 19716, USA}

\author[0009-0005-9366-6163]{Rayta Pradata}
\affiliation{Department of Physics and Astronomy, University of Delaware, Newark, DE 19716, USA}

 \author[0000-0001-7224-6024]{William H. Matthaeus}
\affiliation{Department of Physics and Astronomy, University of Delaware, Newark, DE 19716, USA}

\author[0000-0001-6160-1158]{David J. McComas}
\affiliation{Department of Astrophysical Sciences, Princeton University, Princeton, NJ 08544, USA}

\author[0000-0003-2134-3937]{Eric Christian}
\affiliation{Heliophysics Science Division, NASA Goddard Space Flight Center, Greenbelt, MD 20771, USA}

\author[0000-0002-3737-9283]{Nathan Schwadron}
\affiliation{Los Alamos National Laboratory, PO Box 1663, Los Alamos, NM 87545, USA}

\author[0000-0001-9979-2164]{Matina Gkioulidou}
\affiliation{Johns Hopkins University Applied Physics Laboratory, Laurel, MD 20723, USA}

\author[0000-0002-8111-1444]{Jamie Rankin}
\affiliation{Department of Astrophysical Sciences, Princeton University, Princeton, NJ 08544, USA}

\author[0000-0002-7572-4690]{Timothy S. Horbury}
\affiliation{Department of Physics, Imperial College London, London, SW7 2BW, UK}

\author[0000-0002-4381-3197]{Savvas Raptis}
\affiliation{Johns Hopkins University Applied Physics Laboratory, Laurel, MD 20723, USA}

\author[0000-0002-2425-7818]{Drew L. Turner}
\affiliation{Johns Hopkins University Applied Physics Laboratory, Laurel, MD 20723, USA}

\begin{abstract}

We analyze early Interstellar Mapping and Acceleration Probe (IMAP) magnetic field observations (November 2025 to May 2026) to quantify the statistical properties of helical magnetic structures and their relationship with turbulence intermittency. Intermittent structures, such as current sheets, are identified using the Partial Variance of Increments (PVI) method, and the scale-dependent normalized magnetic helicity is computed over scales $\ell$ corresponding to 0.5--3 correlation lengths ($\lambda_c$). Prior studies have typically applied the PVI-helicity method to individual events; here we extend this technique to the ensemble statistics of a multi-month sample spanning many correlation times. To separate coherent helical structures from random turbulent fluctuations, significance thresholds for helicity are determined using phase-randomized surrogate realizations. We find a clear scale dependence in the occurrence, duration, and waiting time of coherent magnetic helicity, reminiscent of the scale-transfer arguments underlying inverse cascade, selective decay, and Taylor relaxation theories. Event duration and waiting-time distributions are well described by power-law and exponential models over different ranges, consistent with previous findings. In addition, PVI and magnetic helicity are anticorrelated, suggesting that high-PVI regions are preferentially associated with the boundaries of flux ropes, while their interiors tend to exhibit lower PVI and stronger magnetic helicity.

\end{abstract}


\section{Introduction} \label{sec:intro}

Coherent magnetic structures, such as magnetic discontinuities \citep{horbury2001three, zhdankin2012magnetic, greco2016complex, gautam2025characterization}, current sheets \citep{schwartz1985active, li2008identifying, khabarova2015small, greco2018partial, wang2025analysis, Turner2026current}, and flux-tube-like structures \citep{klein2008open, zaqarashvili2014twisted,  zhao2019particle, pecora2019single, zank2021flux}, are frequently observed in the turbulent solar wind over a wide range of spatial scales \citep{brandenburg2011scale, hu2017grad20years, chhiber2020clustering, pecora2021parker}. These structures are important because plasma heating \citep{somov1977current, bowness2013coronal, osman2014magnetic, reeves2019exploring, gautam2026evolution}, particle energization \citep{hess2010power, khabarova2016small, isliker2019particle, nakanotani2020interaction}, and
transport \citep{kaufmann2006magnetic, klein2008open, le2015kinetic, wu2025particle} can become concentrated in their vicinity
and spatially organized by their presence. 

A widely used diagnostic of coherent magnetic structures in spacecraft observations is the Partial Variance of Increments \cite[PVI; see][and references therein]{greco2008intermittent_PVI1,greco2012inhomogeneous,greco2018partial,pecora2019single,pecora2021identification,chhiber2020clustering}. Previous studies \citep{greco2009waiting,chhiber2020clustering} have shown that the temporal distribution of high-PVI structures is nonrandom, with power-law waiting times at shorter separations and approximately exponential behavior at larger separations; their occurrence and clustering also depend on the PVI lag. Magnetic helicity provides complementary information about coherent magnetic structure by characterizing the linkage and winding of magnetic flux \citep{moffatt1978book, matthaeus1982measurement}. Previous studies have also shown that magnetic helicity exhibits systematic variations with spatial scale 
\citep{frisch1975possibility, stribling1994nonlinear} and heliospheric location \citep{bieber1987magnetic,brandenburg2011scale,alberti2022tracking}. These studies mainly describe the broader spatial or spectral organization of helicity. To identify helical structures locally in spacecraft observations, \citet{pecora2021identification} developed a real-space magnetic-helicity method 
based on the general theory provided by \citet{matthaeus1982evaluation_Hm} and combined it with PVI. In this framework, enhanced magnetic helicity indicates the presence of helical flux tubes, whereas strong PVI can identify sharp magnetic gradients near their internal or external boundaries.

Previous studies employing the PVI-helicity flux tube identification method have applied it mainly to specific intervals or events~\citep[see, e.g.,][]{pecora2021parker}. Here, we build on this work using several months of early IMAP observations from November 2025 to May 2026. We examine structures at six different scales, from 0.5 to 3 magnetic correlation lengths, and use phase-randomized surrogate data to set a significance threshold at each scale. This allows us to study how the occurrence, duration, and waiting times of helical structures change with scale, whether positive and negative helicity occur equally often, and how these structures relate to strong magnetic gradients. The statistical characterization of PVI-selected gradients and helical magnetic structures is important for understanding the local topology and dynamics of solar wind turbulence. Our contribution is therefore a statistical description of these properties across a multimonth sample, extending earlier applications of the PVI-helicity method. \vspace{-0.1cm}\\

\section{Data and Methods}

IMAP was launched in September 2025 and carries 10 instruments designed to investigate the interaction of the solar wind with the local interstellar medium and the processes responsible for particle acceleration and transport throughout the heliosphere \citep{Mccomas2025imap}. We use magnetic field and solar wind velocity vector 
measurements from early IMAP observations spanning from November 1, 2025 through May 20, 2026.  Magnetic field data are obtained from the magnetometer (MAG) instrument \citep{horbury2026imap}, while solar-wind velocity measurements are obtained from the Solar Wind and Pickup Ions (SWAPI) instrument \citep{rankin2025solar}. We use magnetic field and velocity cadences of 16~sec and 60~sec, respectively.

We use PVI to identify intermittent magnetic field structures. PVI measures the magnitude of the magnetic field vector increment at a given time lag $\tau$ normalized by the rms amplitude of the increments \citep{greco2018partial}. It thus provides a pointwise measure of the strength of a magnetic field increment relative to the local level of magnetic fluctuations. For a magnetic field $\mathbf{B}(t)$, PVI is defined as
\citep{greco2008intermittent_PVI1,chhiber2020clustering},
\begin{equation}
\mathrm{PVI}(t,\tau) =
\frac{|\Delta\mathbf{B}(t,\tau)|}
{\sqrt{\langle |\Delta\mathbf{B}(t,\tau)|^2\rangle}},
\end{equation}
where
$\Delta\mathbf{B}(t,\tau)
=
\mathbf{B}(t+\tau)-\mathbf{B}(t),$
$t$ is the spacecraft observation time, $\tau$ is the temporal
lag, and $\langle\cdots\rangle$ denotes a local average. We use an averaging window of 6 hours for our analysis. Under Taylor's hypothesis, the six-hour averaging interval corresponds to a spatial extent along the solar wind flow of $L_{\mathrm{avg}}\simeq V_{\mathrm{sw}}\Delta t_{\mathrm{avg}}$, where $\Delta t_{\mathrm{avg}}=6$ hr. For a typical solar wind speed of $400\ \mathrm{km\,s^{-1}}$, this gives $L_{\mathrm{avg}}\simeq8.64\times10^6$ km. Compared with a representative magnetic correlation length of $\lambda_c\sim10^6$ km in our observations, the window spans about nine correlation lengths. This interval defines the local background used for PVI normalization. The PVI values 
are calculated at time lags of $\tau=16$, 60 and 120~s. Large PVI values ($>3$) correspond to fluctuations in the non-Gaussian tails of the increment distribution and are therefore associated with enhanced intermittency \citep{matthaeus2015intermittency, greco2018partial, chhiber2020clustering,pecora2021identification}.

To characterize helical magnetic structures, we estimate the local real-space magnetic helicity following the method of \citet{pecora2021identification, pecora2021parker}. For single-spacecraft measurements, a local estimate of the magnetic helicity at scale $\ell$ is obtained from the antisymmetric part of the magnetic field correlation,

\begin{equation}
H_m(x,\ell)=\int_0^\ell C(x,l)h(l)\,dl ,
\end{equation}
where
\begin{equation}
C(x,l)=\frac{1}{w_0}
\int_{x-w_0/2}^{x+w_0/2}
\left[
b_2(\xi)b_3(\xi+l)-b_3(\xi)b_2(\xi+l)
\right]d\xi,
\end{equation}
and
\begin{equation}
h(l)=\frac{1}{2}
\left[
1+\cos\left(\frac{2\pi l}{w_0}\right)
\right]
\end{equation}
is a Hann window. Here, $b_2$ and $b_3$ are the magnetic-field components perpendicular to the sampling direction $\hat{1}$,
where $\hat{1} =  
\hat{2} \times \hat{3}$,
and $w_0$ is an arbitrary interval of local integration. $x$ denotes the position along the one-dimensional spacecraft sampling direction at which the local magnetic helicity
is evaluated, and $\xi$ is the dummy spatial coordinate used for
integration over the local interval of width $w_0$ centered at $x$. We use $w_0=2\ell$.
The variable $l$ is the spatial lag entering the local correlation,
whereas $\ell$ denotes the maximum spatial scale included in the
helicity integral. We calculate $H_m$ at six different scales,
$\ell/\lambda_c=0.5$, 1.0, 1.5, 2.0, 2.5, and 3.0, where $\lambda_c = V_{\rm sw}\tau_c$ is the magnetic correlation length. The correlation time $\tau_c$ is calculated as the time at which the autocorrelation falls to 1/e value \citep{matthaeus1998dynamical, cuesta2022isotropization, Gautam2024solar}.

The normalized magnetic helicity can be estimated as \citep{pecora2021parker},
\begin{equation}
\widetilde{H}_m(x,\ell)
=
\frac{H_m(x,\ell)}
{\langle\delta b^2\rangle\lambda_c},
\end{equation}
where $\langle\delta b^2\rangle$ is the locally averaged magnetic fluctuation energy. We estimate both $\langle\delta b^2\rangle$ and $\lambda_c$ within each six-hour analysis interval ($\Delta t_{\mathrm{avg}}=6$ hr) because this is an appropriate duration \citep[several correlation times at 1~au,][]{ruiz2014characterization,isaacs2015systematic,cuesta2022isotropization} to attain statistical homogeneity. The analysis is conducted over the first six months of continuous coverage of the IMAP mission at Lagrange point L1.

We determine a scale-dependent threshold for estimated magnetic helicity using phase-randomized surrogate data to distinguish significant helical structures from fluctuations expected in random phase magnetic field data, following approaches similar to those of \cite{sahraoui2008diagnosis,lion2016coherent}. For each 6-hr interval, we generate 100 surrogate magnetic field time series by randomizing the Fourier phases for each component while preserving the Fourier amplitudes. This procedure preserves the power spectrum of the original magnetic field fluctuations while removing the phase coherence associated with coherent structures. The normalized magnetic helicity is then calculated for each surrogate. For each scale $\ell/\lambda_c$, we construct the distribution of $|\widetilde{H}_m|$ from the phase-randomized realizations and define the threshold $H_{\rm th}$ as the 95$^{th}$ percentile of this distribution. Values of $|\widetilde{H}_m|>H_{\rm th}$ are therefore unlikely to arise from random phase fluctuations and are considered as significant helical structures.

We also calculate the waiting time (WT) for PVI events, where consecutive samples satisfying ${\rm PVI}>3$ are treated as a single event. For magnetic helicity, we estimate both the WT and event duration for consecutive samples satisfying $|\widetilde{H}_m|>H_{\rm th}$. The event duration is defined as the time interval over which the event remains continuously above the corresponding threshold. The WT between two successive events is defined as,
${\rm WT}_i =
t_{{\rm start},i+1}-t_{{\rm end},i}$, where $t_{{\rm end},i}$ is the end of the $i^{th}$ event and
$t_{{\rm start},i+1}$ is the beginning of the following event above threshold.

\section{Results}

\begin{figure*}[htpp!]
    \centering
    \includegraphics[width=1\linewidth]{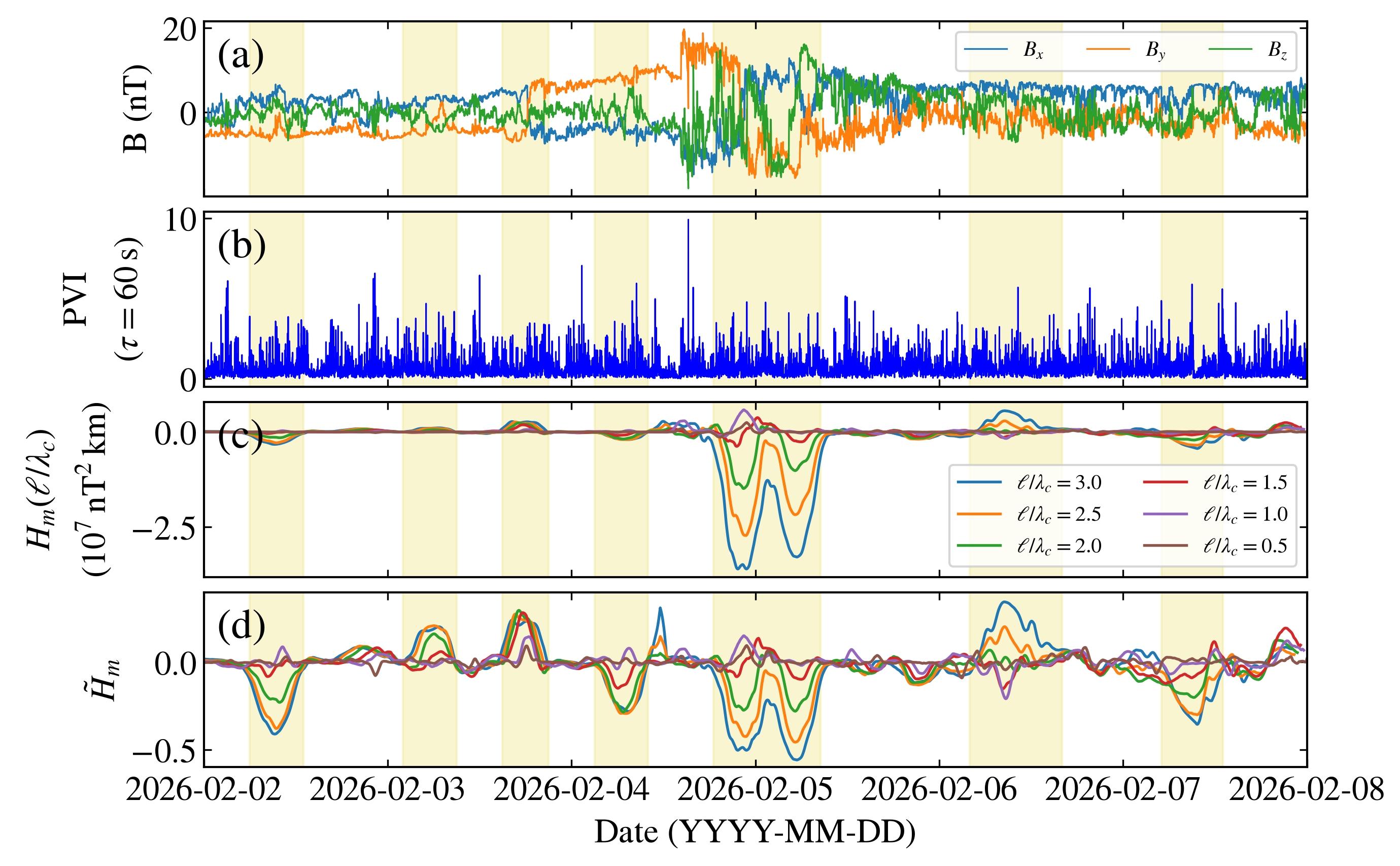}\label{fig:fig1}
    \caption{Example interval showing the identification of helical structures in the solar wind observed by IMAP from 2026 February 2 to 2026 February 8. From top to bottom (a-d), the panels show the magnetic-field components, PVI at a lag of $\tau=60$s, scale-dependent magnetic helicity $H_{m}$, and normalized magnetic helicity $\widetilde{H}_{m}$ for $\ell/\lambda_c=0.5–3.0$. The shaded regions indicate intervals containing helical structures identified against the local phase-randomized threshold.
}
\end{figure*}
\begin{deluxetable*}{ccccccc}[htpp!]
\tablecaption{Occurrence of high-PVI fluctuations at different time lags
\label{tab:pvi_occurrence}}
\tablehead{
\colhead{$\tau$} &
\multicolumn{2}{c}{$\mathrm{PVI}>3$} &
\multicolumn{2}{c}{$\mathrm{PVI}>6$} &
\multicolumn{2}{c}{$\mathrm{PVI}>8$} \\
\colhead{(s)} &
\colhead{$N$} & \colhead{Fraction (\%)} &
\colhead{$N$} & \colhead{Fraction (\%)} &
\colhead{$N$} & \colhead{Fraction (\%)}
}
\startdata
16  & 17,725 & 1.73 & 1,436 & 0.14 & 284 & 0.03 \\
60  & 16,367 & 1.60 &   535 & 0.05 &  57 & 0.01 \\
120 & 13,540 & 1.32 &   222 & 0.02 &   5 & $<0.001$ \\
\enddata
\tablecomments{The fractions represent the percentage of the total
samples at each time lag satisfying the corresponding PVI threshold.}
\end{deluxetable*}
\vspace{-0.8cm}
In Figure 1, we present an example interval illustrating the identification of PVI and magnetic helicity in the IMAP observations from 2026 February 2--8. The panel (a) shows the three magnetic field components, and the panel (b) shows the PVI calculated at a time lag of $\tau=60$ sec. The panel (c) shows the local magnetic helicity $H_m(\ell)$ estimated at six different scales, $\ell/\lambda_c=0.5$, 1.0, 1.5, 2.0, 2.5, and 3.0. Increasing $\ell/\lambda_c$ provides contributions from larger-scale helical structures. The panel (d) shows the corresponding normalized magnetic helicity. The real-space helicity is normalized by the local magnetic field energy. Both positive- and negative-helicity structures are evident throughout the interval, and their characteristic amplitudes and temporal extents vary with $\ell/\lambda_c$. The shaded regions mark representative intervals identified as helical structures with normalized helicity values larger than the local threshold. Several of these regions are also associated with enhanced PVI activity, either at boundaries or in the interior.

\begin{figure}[ht]
    \centering
    \includegraphics[width=\linewidth]{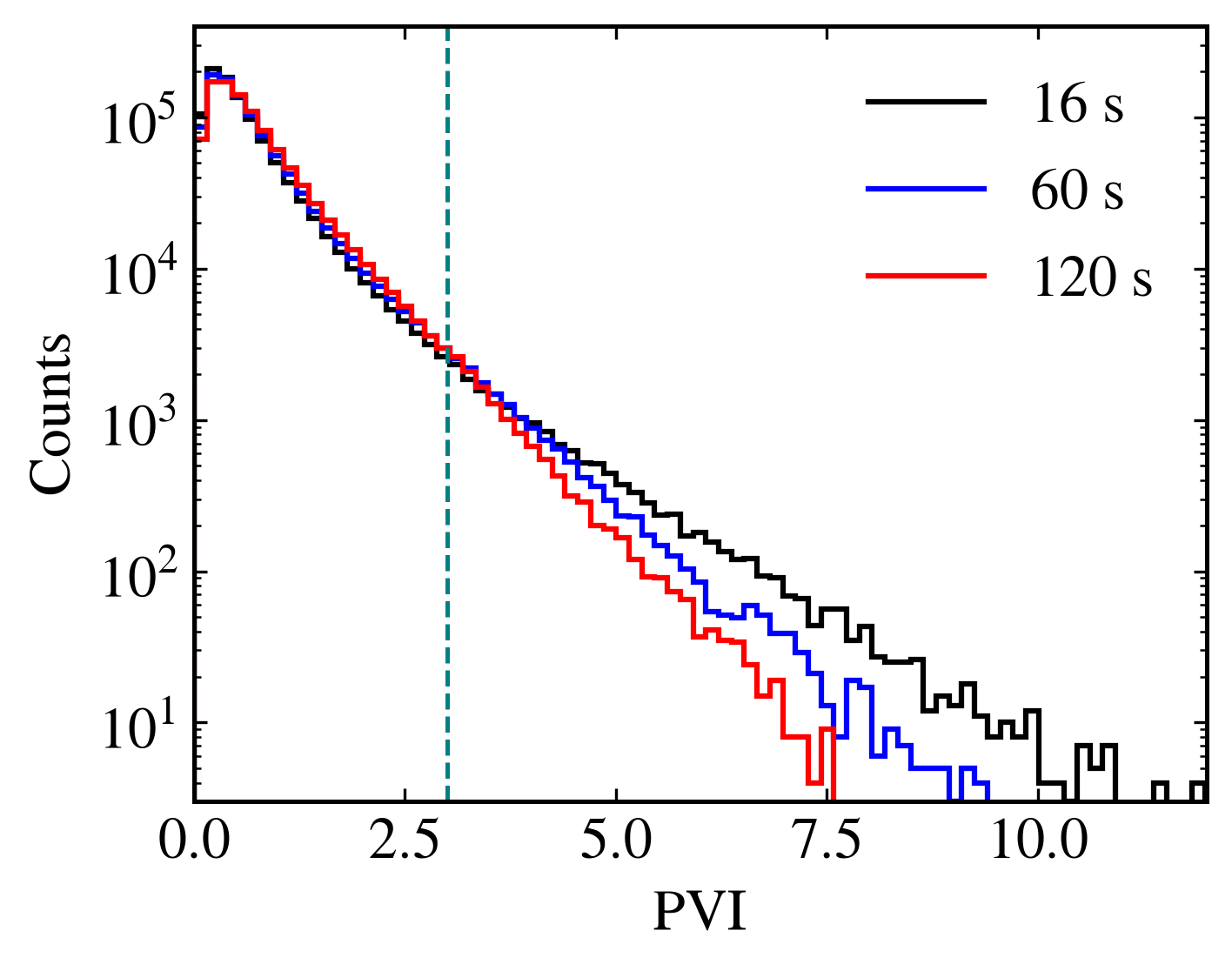}\label{fig:fig2}
    \caption{Distribution of PVI values calculated at time lags of $\tau$ = 16s, 60s, and 120s.}
\end{figure}

Figure~2 shows the distributions of PVI calculated at different time lags of
$\tau=16$, 60, and 120~sec. All three distributions are dominated by
low-PVI fluctuations and are nearly identical over the low-to-moderate
PVI range, but begin to separate beyond $\mathrm{PVI}\sim3$. The
shortest lag shows the broadest high-PVI tail, suggesting a larger
occurrence of strong intermittent fluctuations at smaller scales. Previous studies have
associated $\mathrm{PVI}>3$ with non-Gaussian intermittent structures,
$\mathrm{PVI}>6$ with current sheets, and $\mathrm{PVI}>8$ with
possible magnetic reconnection sites \citep[e.g.,][]{matthaeus2015intermittency,greco2018partial}.
Table~1 summarizes the occurrence of fluctuations above these
thresholds. The number of events decreases gradually with
increasing PVI threshold and time lag, with the strongest fluctuations
occurring mostly at the shortest lag.

\begin{figure}[ht]
    \centering
    \includegraphics[width=\linewidth]{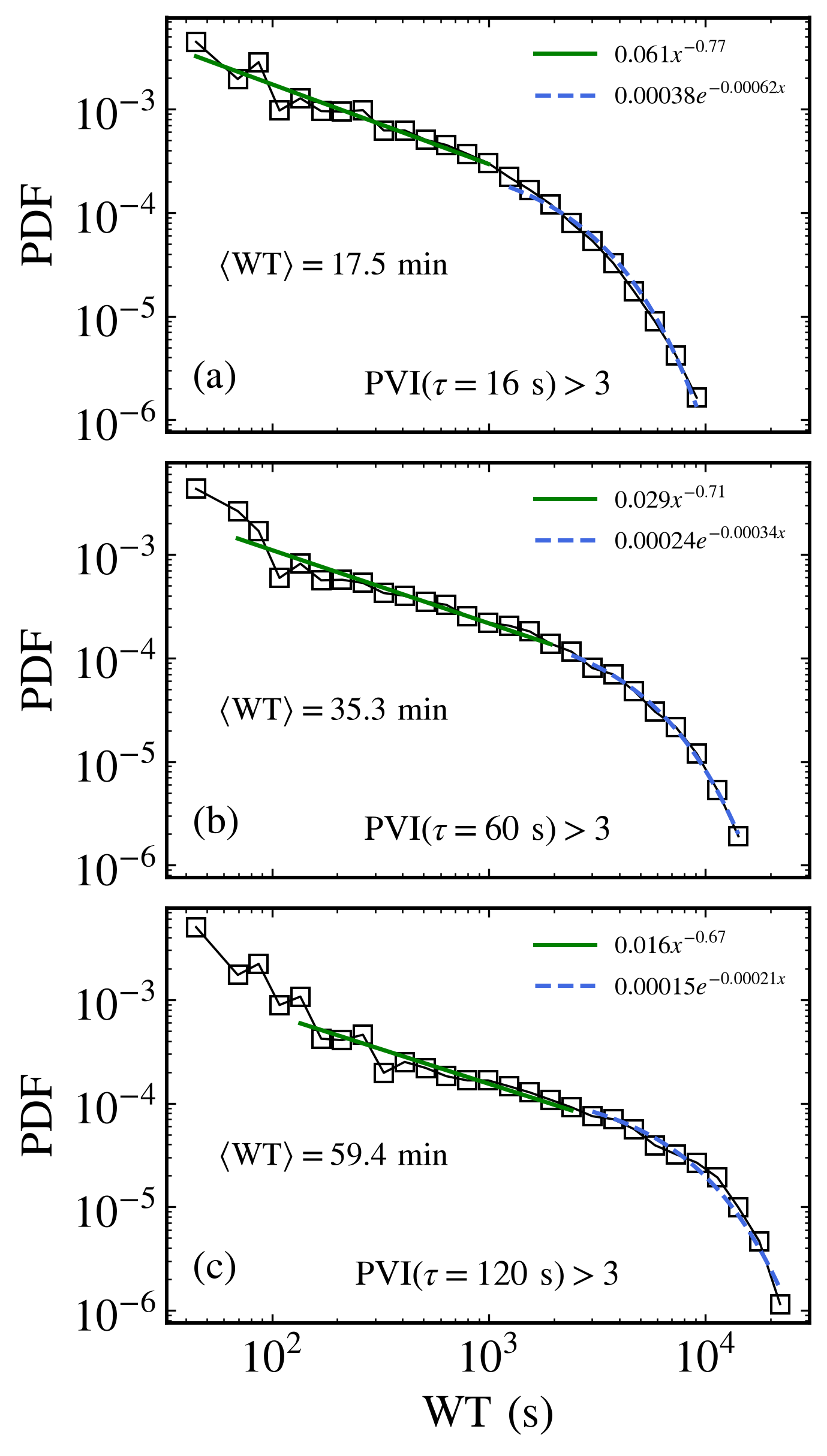}\label{fig:fig3}
    \caption{Probability density functions of waiting times (WTs) between successive PVI events with $\mathrm{PVI}>3$, calculated at time lags of $\tau$ =16s (a), 60s (b), and 120s (c). The green solid lines show power-law fits to the shorter waiting-time regime, while the blue dashed lines show exponential fits to the longer waiting-time tail. The corresponding mean waiting time $\langle \mathrm{WT}\rangle$ is shown in each panel.
 }
\end{figure}

In Figure~3(a-c), we present the probability density functions (PDFs) of WT between successive ${\rm PVI}>3$ events for $\tau=16$, 60, and 120~sec, respectively. For all three lags, the distributions exhibit a power-law behavior at shorter WT followed by an exponential decay at longer WT. This behavior suggests that closely spaced intermittent structures are temporally correlated and clustered, while structures separated by longer intervals are more randomly distributed, consistent with a Poisson noise process. The mean WT, which is measured as the first moment of the distribution, increases from $\langle{\rm WT}\rangle=17.5$~min at $\tau=16$~s to 35.3~min at 60~sec and 59.4~min at 120~sec, showing that strong intermittent events become less frequent with increasing lags.

\begin{figure}[htpp!]
    \centering
    \includegraphics[width=\linewidth]{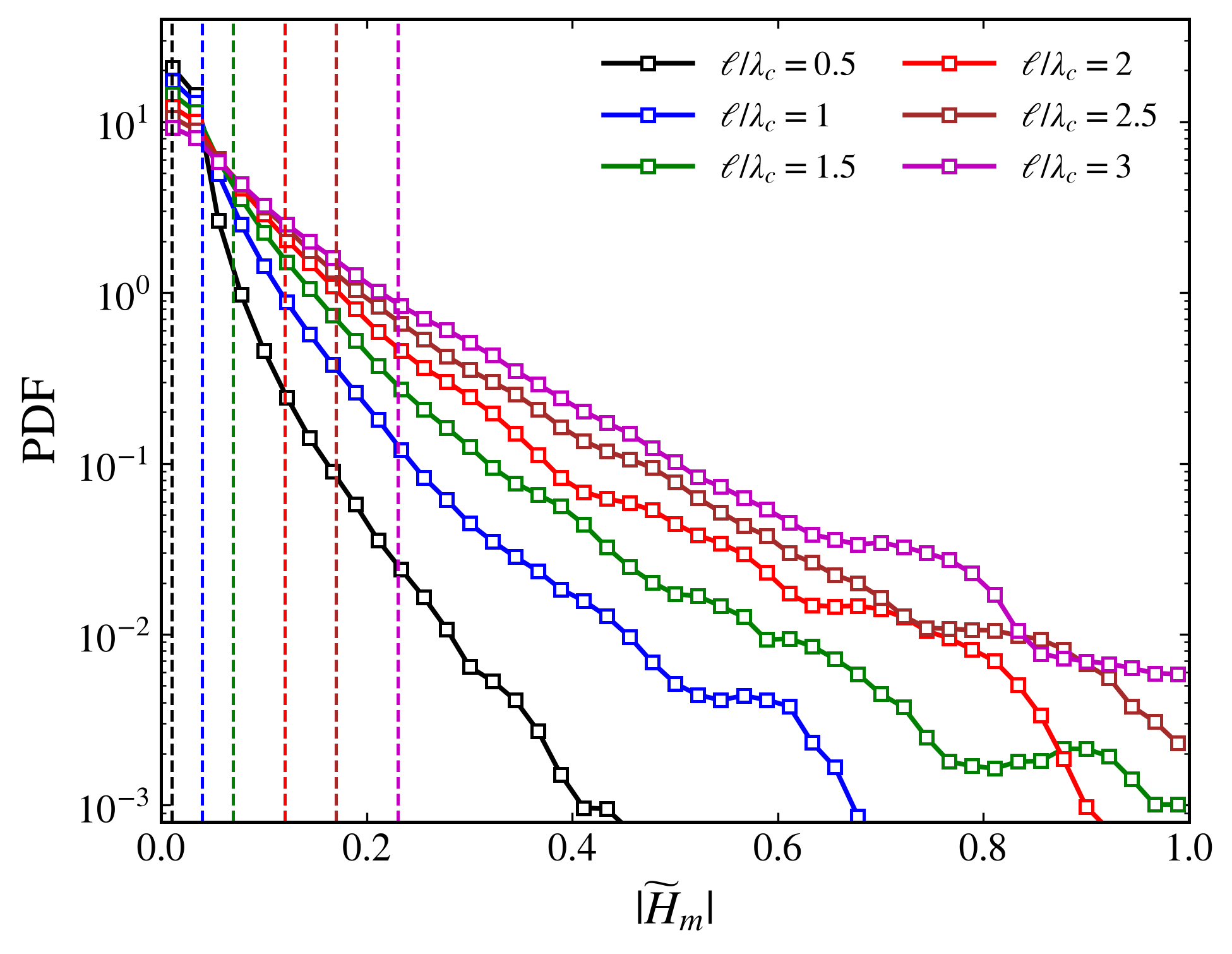}\label{fig:fig4}
    \caption{PDFs of the absolute normalized magnetic helicity $|\widetilde{H}_{m}|$ for scales $\ell/\lambda_c=0.5$--$3.0$. The vertical dashed lines indicate the corresponding scale-dependent threshold values used to identify significant helical structures.}
\end{figure}

\begin{figure}[htpp!]
    \centering
    \includegraphics[width=\linewidth]{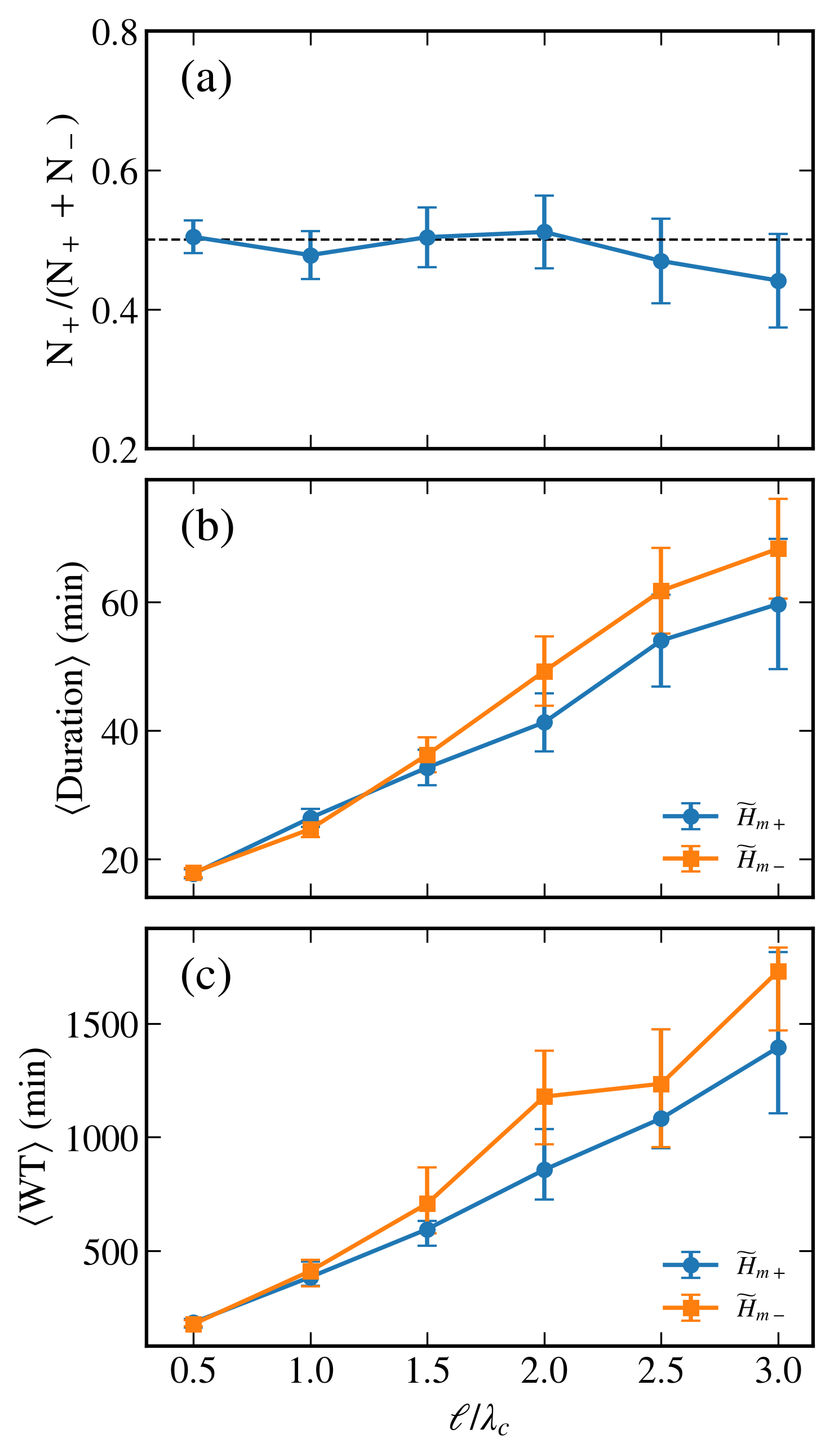}\label{fig:fig5}
    \caption{(a) Fraction of positive-helicity structures, $N_{+}/(N_{+}+N_{-})$, as a function of $\ell/\lambda_c$. Dashed line is the fraction corresponding to 50\%. (b) Mean duration, $\langle \mathrm{Duration}\rangle$, of positive ($\widetilde{H}_{m+}$) and negative ($\widetilde{H}_{m-}$) helicity structures. (c) Mean waiting time, $\langle \mathrm{WT}\rangle$, between successive positive and negative helicity structures. Error bars indicate the corresponding standard deviation.}
\end{figure}

Figure~4 shows the PDFs of the absolute
normalized magnetic helicity $|\widetilde{H}_m|$ for
$\ell/\lambda_c=0.5$--3.0. At the smallest scale ($\ell/\lambda_c=0.5$), the distribution is strongly concentrated at low
values of $|\widetilde{H}_m|$ and decreases rapidly toward larger
helicity. As $\ell/\lambda_c$ increases, the distributions become broader and develop more extended tails toward large
$|\widetilde{H}_m|$. Thus, moderate and large normalized helicity values
become gradually more common at larger scales. The vertical dashed lines indicate the scale-dependent thresholds obtained from the phase-randomized surrogate analysis. The threshold values also increase from 0.011 at
$\ell/\lambda_c=0.5$ to 0.037, 0.075, 0.123, 0.174, and 0.226 at
$\ell/\lambda_c=1$, 1.5, 2, 2.5, and 3, respectively. This increase
indicates the scale dependence of the helicity fluctuations expected
from the corresponding random-phase surrogate data and 
motivates the use of a
scale-dependent threshold rather than a single fixed value. 

The scale dependence evident in the PDFs is also reflected in the
helicity statistics as shown in Table 2. The mean value of $|\widetilde{H}_m|$ increases from 0.016 at $\ell/\lambda_c=0.5$ to 0.102 at
$\ell/\lambda_c=3$. The occurrence of moderately and strongly
helical fluctuations also increases significantly with scale. For
example, the fraction of samples with $|\widetilde{H}_m|>0.25$
increases from 0.11\% at $\ell/\lambda_c=0.5$ to 9.69\% at
$\ell/\lambda_c=3$, a factor of almost 100. 
These results demonstrate that large normalized-helicity
fluctuations become gradually more common with increasing spatial
scale. When the scale-dependent significance thresholds are applied, however, the fraction of samples identified as significant helical structures
decreases with increasing scale. The above-threshold fraction decreases
from 37.85\% at $\ell/\lambda_c=0.5$ to 11.64\% at
$\ell/\lambda_c=3.0$, while the mean $|\widetilde{H}_m|$ of the
above-threshold samples increases from 0.034 to 0.377. Thus, although
large values of normalized helicity become more prevalent at larger
scales, the corresponding significance criterion also becomes more
restrictive because the surrogate-derived threshold increases with
scale.

\begin{deluxetable*}{cccccccc}
\tablecaption{Scale-dependent statistics of normalized magnetic helicity
\label{tab:helicity_stats}}
\tablehead{
\colhead{$\ell/\lambda_c$} &
\colhead{$H_{\rm th}$} &
\colhead{$\langle |\widetilde{H}_m| \rangle$} &
\colhead{$f_{0.25}$ (\%)} &
\colhead{$f_{0.50}$ (\%)} &
\colhead{$f_{0.75}$ (\%)} &
\colhead{$\langle |\widetilde{H}_m| \rangle_{\rm th}$} &
\colhead{$f_{\rm th}$ (\%)}
}
\startdata
0.5 & 0.011 & 0.016 & 0.11 & 0.01 & 0.00 & 0.034 & 37.85 \\
1.0 & 0.037 & 0.033 & 0.78 & 0.08 & 0.02 & 0.087 & 27.32 \\
1.5 & 0.075 & 0.050 & 2.16 & 0.28 & 0.05 & 0.152 & 20.51 \\
2.0 & 0.123 & 0.068 & 4.24 & 0.67 & 0.14 & 0.227 & 15.80 \\
2.5 & 0.174 & 0.085 & 6.75 & 1.09 & 0.28 & 0.306 & 13.01 \\
3.0 & 0.226 & 0.102 & 9.69 & 1.70 & 0.46 & 0.377 & 11.64 \\
\enddata
\tablecomments{
$H_{\rm th}$ is the scale-dependent threshold. $\langle|\widetilde{H}_m|\rangle$ denotes the mean absolute normalized
magnetic helicity calculated using all samples, whereas
$\langle|\widetilde{H}_m|\rangle_{\rm th}$ denotes the mean calculated
only for samples satisfying $|\widetilde{H}_m|>H_{\rm th}$. $f_{\rm th}$ is the corresponding
fraction of samples above the threshold. The quantities $f_{0.25}$,
$f_{0.50}$, and $f_{0.75}$ denote the percentages of samples satisfying
$|\widetilde{H}_m|>0.25$, 0.50, and 0.75, respectively.
}
\end{deluxetable*} \vspace{-0.8cm}
Figure~5 compares the statistical properties of positive and
negative helicity structures as a function of $\ell/\lambda_c$.
Panel (a) shows the fraction of positive helicity events,
$N_{+}/(N_{+}+N_{-})$, with the horizontal dashed line marking 0.5 (equal occurrence of positive and negative helicity
events). The fraction remains close to 0.5 across all scales, indicating
no strong preference for either helicity sign, although a modest
decrease in the positive helicity fraction is observed at the largest
scales. Panels (b) and (c) compare the mean event durations and WTs, respectively. The mean durations of both helicity signs increase
with scale and are comparable at smaller scales, whereas
negative helicity structures tend to have slightly longer durations for
$\ell/\lambda_c\gtrsim2$. The mean WT also increases
with scale, with negative helicity structures generally showing longer
WT than positive helicity structures at the larger scales.

\begin{figure*}[htpp!]
    \centering
    \includegraphics[width=\linewidth]{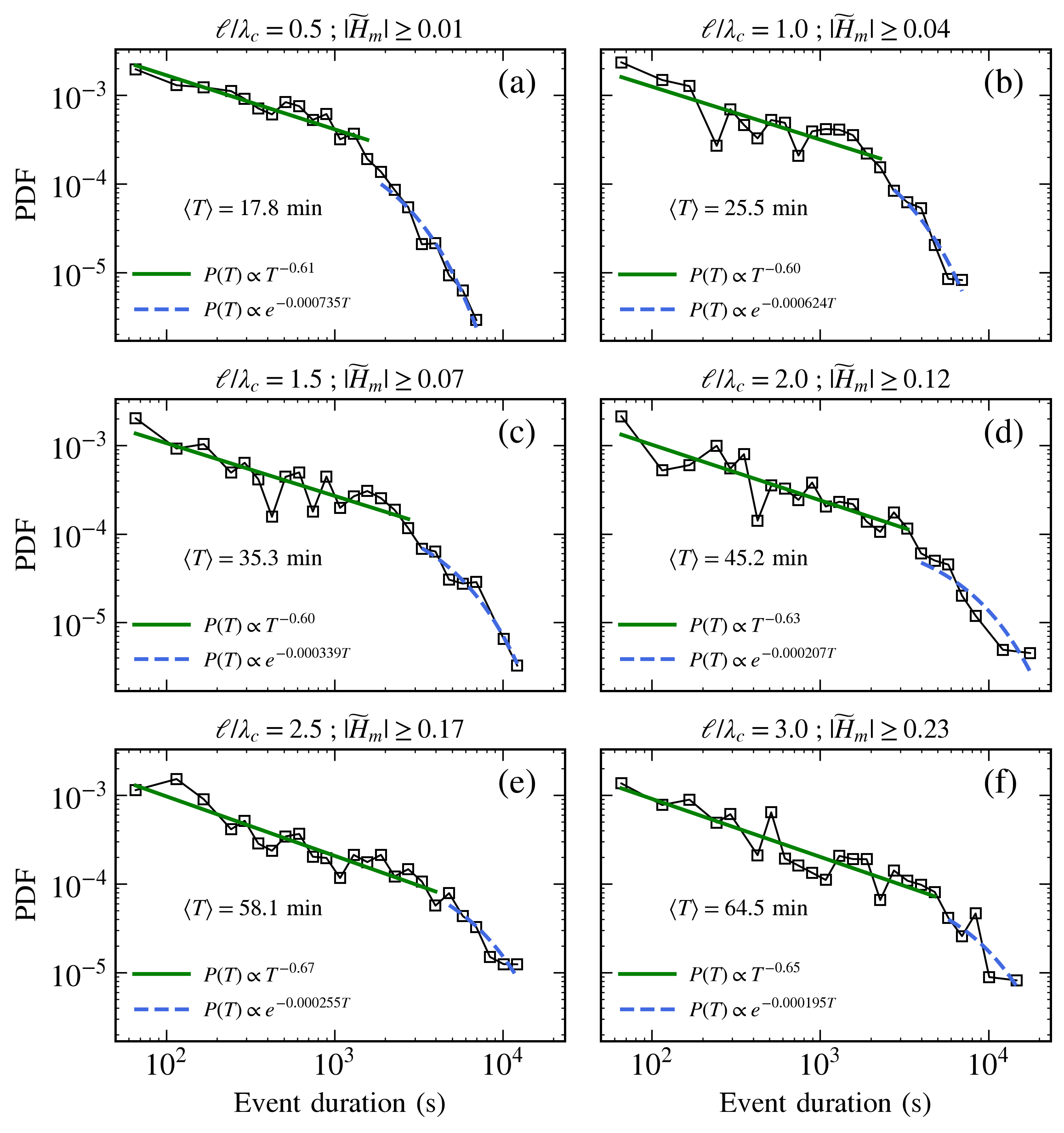}\label{fig:fig6}
    \caption{PDFs of the durations of helical structures at scales $\ell/\lambda_c=0.5$--$3.0$ (a-f). The green solid lines show power-law fits to the shorter-duration regime, while the blue dashed lines show exponential fits to the longer-duration tail. The corresponding mean duration is also provided in each panel.}
\end{figure*}

\begin{figure*}[ht]
    \centering
    \includegraphics[width=\linewidth]{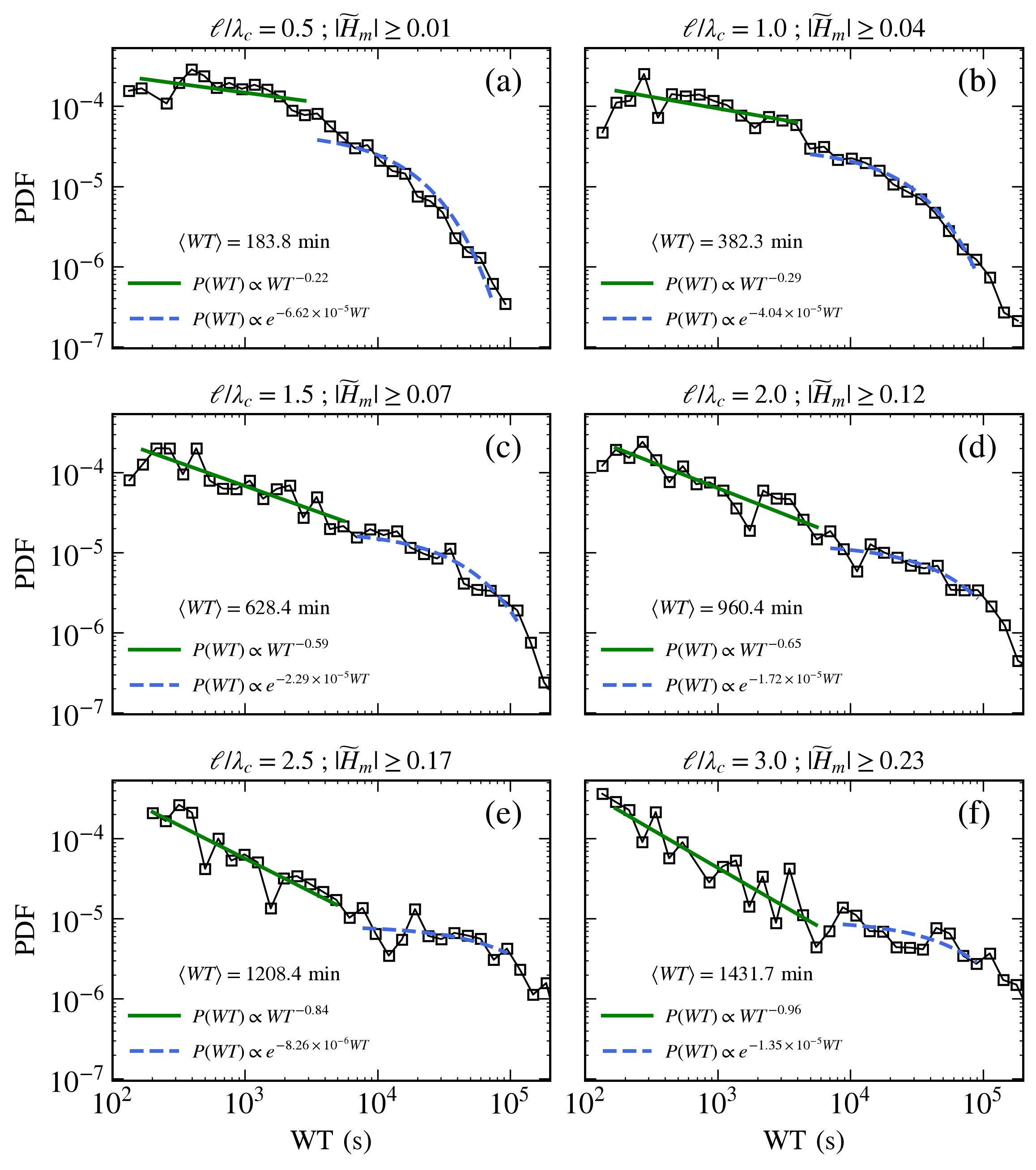}\label{fig:fig7}
    \caption{Same as Figure~6, but showing the PDFs of the waiting times between successive helical structures.}
\end{figure*} 

Figure~6(a-f) shows the PDFs of the durations of helical structures at different scales ($\ell/\lambda_c$ = 0.5--3). At all
scales, the distributions exhibit a power-law behavior
at shorter durations, followed by a more rapid, approximately
exponential decrease at longer durations. The fitted power-law indices
show relatively little variation with scale, ranging from about
$-0.60$ to $-0.67$. In contrast, the duration distributions extend
progressively toward longer times as $\ell/\lambda_c$ increases.
Accordingly, the mean event duration increases from
$\langle T\rangle=17.8$~min at $\ell/\lambda_c=0.5$ to
64.5~min at $\ell/\lambda_c=3.0$. This scale dependence indicates that
helical structures identified at larger spatial scales generally
persist for longer periods. In a similar format, Fig.~7 shows the PDFs of the WTs between successive helical structures. The WT distributions also exhibit similar fitting trends to the event duration distributions.
The short WT power-law slope becomes gradually steeper with
increasing $\ell/\lambda_c$, changing from approximately $-0.22$ at
$\ell/\lambda_c=0.5$ to $-0.96$ at $\ell/\lambda_c=3.0$. The
distributions also extend toward longer waiting times as
the scale increases. Accordingly, the mean waiting time increases from
$\langle{\rm WT}\rangle=183.8$~min at $\ell/\lambda_c=0.5$ to
1431.7~min at $\ell/\lambda_c=3.0$. This increasing trend indicates
that helical structures at larger scales are more widely
separated in time and therefore occur less frequently. We note that although the duration and WT PDFs exhibit qualitatively similar behavior, our analysis of their joint probability distribution (not shown here) reveals no significant relationship between the duration of a helical structure and the WT between successive structures. Thus, longer-lived structures are not necessarily followed by longer or shorter waiting times.

\begin{figure*}[htpp!]
    \centering
    \includegraphics[width=0.6\linewidth]{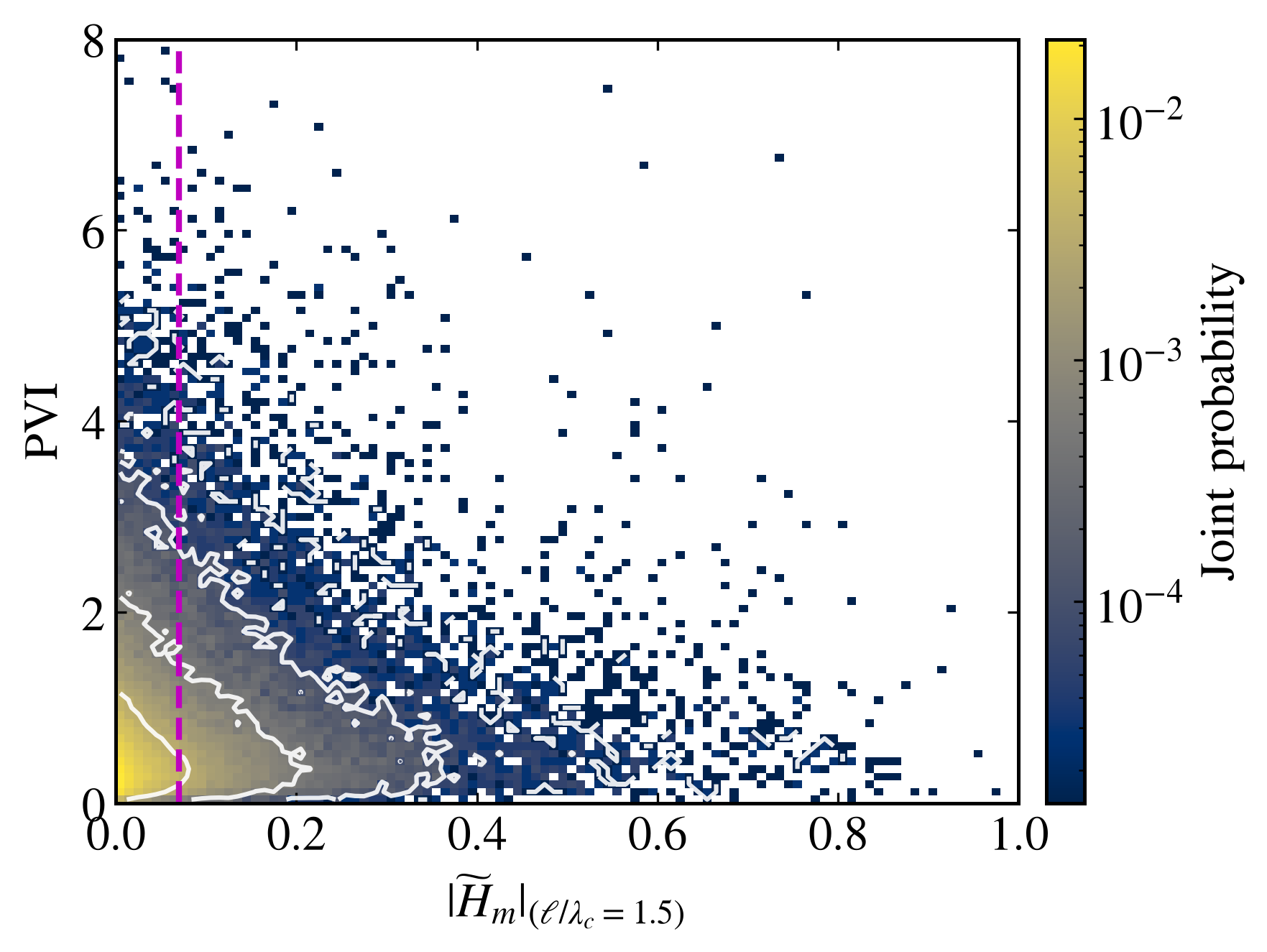}\label{fig:fig8}
    \caption{Joint probability distribution of PVI and $|\widetilde{H}_m|$ at $\ell/\lambda_c=1.5$. White contours indicate levels of constant joint probability. The vertical magenta dashed line marks the helicity threshold used to identify significant helical structures at $\ell/\lambda_c=1.5$.}
\end{figure*}

We further examine the PVI-helicity relationship. In this framework, extrema of the local magnetic helicity tend to
identify the cores of helical structures, whereas enhanced PVI is
preferentially associated with sharp magnetic gradients near their
boundaries. Thus, large values of PVI and $|\widetilde{H}_m|$ are not
necessarily expected to occur simultaneously, and in fact their occurrences are more likely to be anti-correlated. 
We consider PVI at $\tau=60$~sec and $|\widetilde{H}_m|$ at
$\ell/\lambda_c=1.5$ for this analysis.  Figure~8 shows the joint probability distribution of PVI and $|\widetilde{H}_m|$. The vertical dashed line marks the helicity threshold $H_{\rm th}=0.075$. The distribution is strongly concentrated at
low PVI and low $|\widetilde{H}_m|$, while the probability decreases
toward larger values of either quantity. Large PVI fluctuations occur
mostly at relatively small $|\widetilde{H}_m|$, whereas
samples with enhanced $|\widetilde{H}_m|$ are concentrated mainly at
lower PVI. This behavior is qualitatively consistent with the picture
of \citet{pecora2021identification}, in which helicity maxima occur
within relatively PVI-quiet regions and enhanced PVI marks nearby
boundaries of the helical structures.\vspace{0.2cm}

\section{Discussion and Conclusions}

\begin{figure*}[htpp!]
    \centering
    \includegraphics[width=\linewidth]{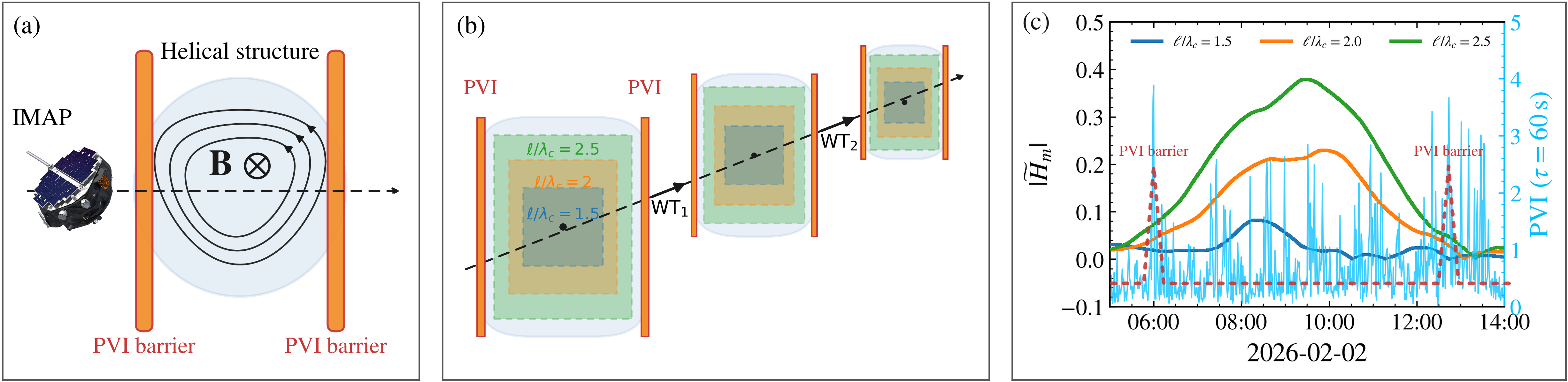}\label{fig:fig9}
    \caption{Summary illustration of the relationship between magnetic-helicity structures and intermittent PVI events: (a) illustration of a helical magnetic structure bounded by enhanced-PVI regions, (b) schematic representation of scale-dependent helical structures and the WTs between successive structures, and (c) a representative IMAP interval of $|\widetilde{H}_m|$ at different spatial scales together with PVI calculated at a lag of 60~s.}
\end{figure*}

We used early-mission
IMAP observations to examine coherent magnetic
structures from two complementary perspectives, (i) strong magnetic field
gradients identified by PVI, and (ii) locally helical structures identified
from the real-space magnetic helicity. A consistent picture emerging
from the analysis is that both populations are strongly organized by
scale. The high-PVI tail becomes less populated as the PVI lag
increases, while the characteristic strength, duration, and separation
of helical structures change systematically with $\ell/\lambda_c$.
An important aspect of the helicity analysis is that the background
level itself is scale dependent. Although large values of
$|\widetilde{H}_m|$ become more common toward larger scales, the
surrogate-derived significance threshold also increases with scale.
Consequently, a single fixed helicity threshold would not provide an
equivalent significance criterion at different $\ell/\lambda_c$ (as shown in Table 2). The main results of the present analysis are summarized
schematically in Figure~9. Panel (a) illustrates
the physical picture in which a helical magnetic structure is
bounded by sharp magnetic field gradients identified by enhanced
PVI. The panel (b) summarizes the scale-dependent organization of the
helical structures and the WTs between successive
structures. Panel (c) shows a representative IMAP interval
in which the helicity signatures at different spatial scales are
viewed together with the PVI.

The observed PVI WT distributions are consistent with previous results near Earth and closer to the Sun \citep{greco2009waiting,chhiber2020clustering}, showing power-law behavior at shorter separations and an approximately exponential tail at larger separations. The estimated mean WTs are comparable to near 1~au timescales \citep[see][and references therein]{greco2009waiting} and longer than those reported during Parker Solar Probe's (PSP) first perihelion \citep{chhiber2020clustering}. The shorter WTs observed by PSP may reflect the smaller turbulence correlation scales and more rapidly varying magnetic structures in the young solar wind, which evolve and become more widely separated as the solar wind expands toward 1 au.

The duration statistics of magnetic helicity provide a connection with previous flux-rope studies. Small-scale magnetic flux ropes identified at 1~au have been found to span durations from several minutes to several
hours and to exhibit scale-dependent occurrence statistics
\citep{zheng2018observational,farooki2024closer}. More recently,
\citet{lynch2023s} applied the $H_m$--PVI technique to
slow-to-moderate solar wind at 1~au and reported mean coherent structure
durations of approximately one hour. This is comparable to the mean
duration found here at the largest scales, where
$\langle T\rangle$ approaches one hour. 

The present analysis further
shows that the characteristic duration of helical structures at scale $\ell$
changes with
$\ell/\lambda_c$, while the short-duration power-law index remains
relatively similar across the scales considered. This suggests that
the characteristic size of the structures changes significantly with
scale even though the statistical form of the short-duration
population is comparatively similar. The WT statistics further show that the temporal spacing of the helicity-selected structures also depends strongly on scale. This statistical behavior is represented conceptually in
Figure~9(b), where structures identified at
different spatial scales occupy different temporal extents and
successive helical structures are separated by finite WTs. The increasing characteristic duration and WT
with $\ell/\lambda_c$ therefore indicate a systematic
scale-dependent organization of the coherent magnetic structures. The power-law behavior at shorter WT followed by
an exponential tail resembles the statistical organization previously
reported for magnetic discontinuities
\citep{greco2009waiting,chhiber2020clustering}. An additional possible interpretation of these ensemble statistics
is suggested by the superposition framework of \citet{schwadron2010superposition}.
In that framework, an observed distribution may arise from the
superposition of individual stochastic realizations characterized
by different local parameters or occurrence rates. In particular,
a superposition of Poisson-like processes having different rates
can produce power-law-like WT statistics even when the
individual realizations have exponential WT distributions.
The solar wind intervals sampled here likely represent a range of
local turbulence and coherent-structure conditions, and therefore
such a superposition may contribute to the observed duration and
WT distributions. Establishing a quantitative connection,
however, would require characterizing the distributions of the
individual realizations and relating them to the corresponding
ensemble statistics, which is beyond the scope of the present study. The joint PVI-helicity statistics are also consistent with the
complementary interpretation proposed by
\citet{pecora2021identification}. As summarized in Figure~9(a), enhanced helicity
mainly characterizes the interior of the coherent structure,
whereas large PVI values preferentially mark sharp gradients near
its boundaries. The representative IMAP interval in
Figure~9(c) provides an observational example of
this organization, with scale-dependent helicity enhancements
occurring predominantly within comparatively PVI-quiet intervals
and strong PVI enhancements appearing near the boundaries of the
helical structures.

We find nearly comparable occurrence rates of positive and
negative helicity structures over the scales examined. This should not
be interpreted as contradicting earlier observations of hemispheric
helicity preferences \citep{bieber1987magnetic,
brandenburg2011scale}. Those studies concern still larger-scale averaged helicity organized relative to heliographic
hemisphere, whereas the present analysis characterizes local helical
structures sampled near 1 au. Interestingly, although the
occurrence of the two signs remains approximately balanced, differences
in their durations and waiting times become more evident at larger
scales. Whether this asymmetry is persistent or depends on solar wind
type, magnetic sector, or 
position relative to 
the heliospheric current sheet
will require a larger IMAP dataset
that better samples 
solar magnetic polarities.

By extending the PVI-helicity method to several months of early IMAP observations, we show how the statistical properties of helical structures vary with scale when identified using surrogate-based significance thresholds. As the dataset grows, the same analysis can be applied to different solar wind types, magnetic sector, heliospheric current sheet proximity, transient activity, heliocentric distance, and solar activity phase to better understand the scale-dependent properties reported
here. The broader IMAP instrument suite will also permit the
identified helical structures and their PVI-rich boundaries, shown
schematically in Figure~9, to be related to solar wind plasma properties, energetic particle populations, and composition. Such studies are particularly relevant in view of previous evidence that helical flux tubes can act as particle-transport barriers~\citep{seripienlert2010dropouts,pecora2021identification}.\vspace{-0.38cm}\\
\section{Acknowledgments}
This research was partially supported by the NASA IMAP project 80GSFC19C0027 at Princeton, under a subcontract SUB0000317 at the University of Delaware
and in part by the PSP/IS\(\odot\)IS project through subcontract SUB0000165 and from Princeton, as well as
NSF SHINE grant AGS- 2501387,
and by NSF SHINE 2501387 at the University of Delaware. F.P. was partially supported by Agenzia Spaziale Italiana (ASI) in the framework of the project “Radial Evolution of large- and KInetic-scale Processes in the Expanding Solar wind (REKIPES)” C83C25000900005. J.G. was supported by the Delaware NASA Space Grant program grant number 80NSSC20M0045 at the University of Delaware. R.B. acknowledges support from NSF FDSS Award No. AGS-2347952. IMAP magnetometer operations are funded by the UK Space Agency (grant UKRI2494). Tim Horbury is supported by STFC grant UKRI1202.

\bibliography{references, refences_ciccio}

\end{document}